
\overfullrule0pt
\input phyzzx.tex
%
%
\newbox\hdbox%
\newcount\hdrows%
\newcount\multispancount%
\newcount\ncase%
\newcount\ncols
\newcount\nrows%
\newcount\nspan%
\newcount\ntemp%
\newdimen\hdsize%
\newdimen\newhdsize%
\newdimen\parasize%
\newdimen\thicksize%
\newdimen\thinsize%
\newdimen\tablewidth%
\newif\ifcentertables%
\newif\ifendsize%
\newif\iffirstrow%
\newif\iftableinfo%
\newtoks\dbt%
\newtoks\hdtks%
\newtoks\savetks%
\newtoks\tableLETtokens%
\newtoks\tabletokens%
\newtoks\widthspec%
%
%
\immediate\write15{%
CP SMSG GJMSINK TEXTABLE -----> TABLE MACROS LOADED, JOB = \jobname%
}%
%
%
\tableinfotrue%
\catcode`\@=11
\def\tstrut{\vrule height16pt depth6pt width0pt}%
\def\|{|}
\def\tablerule{\noalign{\hrule height\thinsize depth0pt}}%
\thicksize=1.5pt
\thinsize=0.6pt
\def\thickrule{\noalign{\hrule height\thicksize depth0pt}}%
\def\ctr#1{\hfil\ #1\hfil}%
%
%
%
\tablewidth=-\maxdimen%
\def\tabskipglue{0pt plus 1fil minus 1fil}%
%
%
\centertablestrue%
%
%
%
%
\parasize=4in%
\gdef\ARGS{########}
\gdef\headerARGS{####}
\def\@mpersand{&}
{\catcode`\|=13
\gdef\letbarzero{\let|0}
\gdef\letbartab{\def|{&&}}
}
{\def\ampskip{&\omit\hfil&}
\catcode`\&=13
\let&0
\xdef\letampskip{\def&{\ampskip}}
}
\def\begintable{
   \begingroup%
   \catcode`\|=13\letbartab%
   \catcode`\&=13\letampskip%
   \def\multispan##1{
      \omit \mscount##1%
      \multiply\mscount\tw@\advance\mscount\m@ne%
      \loop\ifnum\mscount>\@ne \sp@n\repeat%
   }
   \def\|{%
      &\omit\widevline&%
   }%
   \ruledtable
}
\long\def\ruledtable#1\endtable{%
%
%
%
   \offinterlineskip
   \tabskip 0pt
   \def\widevline{\vrule width\thicksize}
   \def\endrow{\@mpersand\omit\hfil\crnorm\@mpersand}%
   \def\crthick{\@mpersand\crnorm\thickrule\@mpersand}%
   \def\crnorule{\@mpersand\crnorm\@mpersand}%
   \let\nr=\crnorule
   \def\endtable{\@mpersand\crnorm\thickrule}%
   \let\crnorm=\cr
%
%
   \edef\cr{\@mpersand\crnorm\tablerule\@mpersand}%
   \the\tableLETtokens
%
%
   \tabletokens={&#1}
%
%
   \countROWS\tabletokens\into\nrows%
   \countCOLS\tabletokens\into\ncols%
%
%
   \advance\ncols by -1%
   \divide\ncols by 2%
   \advance\nrows by 1%
%
%
   \iftableinfo %
      \immediate\write16{[Nrows=\the\nrows, Ncols=\the\ncols]}%
   \fi%
%
%
   \ifcentertables
      \line{
      \hss
   \else %
      \hbox{%
   \fi
      \vbox{%
         \makePREAMBLE{\the\ncols}
         \edef\next{\preamble}
         \let\preamble=\next
         \makeTABLE{\preamble}{\tabletokens}
      }
      \ifcentertables \hss}\else }\fi
   \endgroup
   \tablewidth=-\maxdimen
}
\def\makeTABLE#1#2{
   {
   \let\ifmath0
   \let\header0
   \let\multispan0
%
%
   \ifdim\tablewidth>-\maxdimen %
 \widthspec=\expandafter{\expandafter t\expandafter o%
 \the\tablewidth}%
   \else %
      \widthspec={}%
   \fi %
   \xdef\next{
      \halign\the\widthspec{%
      #1
      \noalign{\hrule height\thicksize depth0pt}
      \the#2\endtable
%
      }
   }
   }
   \next
}
\def\makePREAMBLE#1{
   \ncols=#1
   \begingroup
   \let\ARGS=0
   \edef\xtp{\widevline\ARGS\tabskip\tabskipglue%
   &\tstrut\ctr{\ARGS}}
   \advance\ncols by -1
   \loop
      \ifnum\ncols>0 %
      \advance\ncols by -1%
      \edef\xtp{\xtp&\vrule width\thinsize\ARGS&\ctr{\ARGS}}%
   \repeat
   \xdef\preamble{\xtp&\widevline\ARGS\tabskip0pt%
   \crnorm}
   \endgroup
}
\def\countROWS#1\into#2{
   \let\countREGISTER=#2%
   \countREGISTER=0%
   \expandafter\ROWcount\the#1\endcount%
}%
\def\ROWcount{%
   \afterassignment\subROWcount\let\next= %
}%
\def\subROWcount{%
   \ifx\next\endcount %
      \let\next=\relax%
   \else%
      \ncase=0%
      \ifx\next\cr %
         \global\advance\countREGISTER by 1%
         \ncase=0%
      \fi%
      \ifx\next\endrow %
         \global\advance\countREGISTER by 1%
         \ncase=0%
      \fi%
      \ifx\next\crthick %
         \global\advance\countREGISTER by 1%
         \ncase=0%
      \fi%
      \ifx\next\crnorule %
         \global\advance\countREGISTER by 1%
         \ncase=0%
      \fi%
      \ifx\next\header %
         \ncase=1%
      \fi%
      \relax%
      \ifcase\ncase %
         \let\next\ROWcount%
      \or %
         \let\next\argROWskip%
      \else %
      \fi%
   \fi%
   \next%
}
\def\counthdROWS#1\into#2{%
\dvr{10}%
   \let\countREGISTER=#2%
   \countREGISTER=0%
\dvr{11}%
\dvr{13}%
   \expandafter\hdROWcount\the#1\endcount%
\dvr{12}%
}%
\def\hdROWcount{%
   \afterassignment\subhdROWcount\let\next= %
}%
\def\subhdROWcount{%
   \ifx\next\endcount %
      \let\next=\relax%
   \else%
      \ncase=0%
      \ifx\next\cr %
         \global\advance\countREGISTER by 1%
         \ncase=0%
      \fi%
      \ifx\next\endrow %
         \global\advance\countREGISTER by 1%
         \ncase=0%
      \fi%
      \ifx\next\crthick %
         \global\advance\countREGISTER by 1%
         \ncase=0%
      \fi%
      \ifx\next\crnorule %
         \global\advance\countREGISTER by 1%
         \ncase=0%
      \fi%
      \ifx\next\header %
         \ncase=1%
      \fi%
\relax%
      \ifcase\ncase %
         \let\next\hdROWcount%
      \or%
         \let\next\arghdROWskip%
      \else %
      \fi%
   \fi%
   \next%
}%
{\catcode`\|=13\letbartab
\gdef\countCOLS#1\into#2{%
   \let\countREGISTER=#2%
   \global\countREGISTER=0%
   \global\multispancount=0%
   \global\firstrowtrue
   \expandafter\COLcount\the#1\endcount%
   \global\advance\countREGISTER by 3%
   \global\advance\countREGISTER by -\multispancount
}%
\gdef\COLcount{%
   \afterassignment\subCOLcount\let\next= %
}%
{\catcode`\&=13%
\gdef\subCOLcount{%
   \ifx\next\endcount %
      \let\next=\relax%
   \else%
      \ncase=0%
      \iffirstrow
         \ifx\next& %
            \global\advance\countREGISTER by 2%
            \ncase=0%
         \fi%
         \ifx\next\span %
            \global\advance\countREGISTER by 1%
            \ncase=0%
         \fi%
         \ifx\next| %
            \global\advance\countREGISTER by 2%
            \ncase=0%
         \fi
         \ifx\next\|
            \global\advance\countREGISTER by 2%
            \ncase=0%
         \fi
         \ifx\next\multispan
            \ncase=1%
            \global\advance\multispancount by 1%
         \fi
         \ifx\next\header
            \ncase=2%
         \fi
         \ifx\next\cr       \global\firstrowfalse \fi
         \ifx\next\endrow   \global\firstrowfalse \fi
         \ifx\next\crthick  \global\firstrowfalse \fi
         \ifx\next\crnorule \global\firstrowfalse \fi
      \fi
\relax
      \ifcase\ncase %
         \let\next\COLcount%
      \or %
         \let\next\spancount%
      \or %
         \let\next\argCOLskip%
      \else %
      \fi %
   \fi%
   \next%
}%
\gdef\argROWskip#1{%
   \let\next\ROWcount \next%
}
\gdef\arghdROWskip#1{%
   \let\next\ROWcount \next%
}
\gdef\argCOLskip#1{%
   \let\next\COLcount \next%
}
}
}
\def\spancount#1{
   \nspan=#1\multiply\nspan by 2\advance\nspan by -1%
   \global\advance \countREGISTER by \nspan
   \let\next\COLcount \next}%
\def\dvr#1{\relax}%
\def\header#1{%
\dvr{1}{\let\cr=\@mpersand%
\hdtks={#1}%
\counthdROWS\hdtks\into\hdrows%
\advance\hdrows by 1%
\ifnum\hdrows=0 \hdrows=1 \fi%
\dvr{5}\makehdPREAMBLE{\the\hdrows}%
\dvr{6}\getHDdimen{#1}%
{\parindent=0pt\hsize=\hdsize{\let\ifmath0%
\xdef\next{\valign{\headerpreamble #1\crnorm}}}\dvr{7}\next\dvr{8}%
}%
}\dvr{2}}
\def\makehdPREAMBLE#1{
\dvr{3}%
\hdrows=#1
{
\let\headerARGS=0%
\let\cr=\crnorm%
\edef\xtp{\vfil\hfil\hbox{\headerARGS}\hfil\vfil}%
\advance\hdrows by -1
\loop
\ifnum\hdrows>0%
\advance\hdrows by -1%
\edef\xtp{\xtp&\vfil\hfil\hbox{\headerARGS}\hfil\vfil}%
\repeat%
\xdef\headerpreamble{\xtp\crcr}%
}
\dvr{4}}
\def\getHDdimen#1{%
\hdsize=0pt%
\getsize#1\cr\end\cr%
}
\def\getsize#1\cr{%
\endsizefalse\savetks={#1}%
\expandafter\lookend\the\savetks\cr%
\relax \ifendsize \let\next\relax \else%
\setbox\hdbox=\hbox{#1}\newhdsize=1.0\wd\hdbox%
\ifdim\newhdsize>\hdsize \hdsize=\newhdsize \fi%
\let\next\getsize \fi%
\next%
}%
\def\lookend{\afterassignment\sublookend\let\looknext= }%
\def\sublookend{\relax%
\ifx\looknext\cr %
\let\looknext\relax \else %
   \relax
   \ifx\looknext\end \global\endsizetrue \fi%
   \let\looknext=\lookend%
    \fi \looknext%
}%
%
%
\def\tablelet#1{%
   \tableLETtokens=\expandafter{\the\tableLETtokens #1}%
}%
\catcode`\@=12
%
\def\ifmath#1{\relax\ifmmode #1\else $#1$\fi}
\def\ZPC#1#2#3{{\sl Z.~Phys.} {\bf C#1}, #2 (#3)}
\def\PTP#1#2#3{{\sl Prog. Theor. Phys.} {\bf #1}, #2 (#3)}
\def\PRL#1#2#3{{\sl Phys. Rev. Lett.} {\bf #1}, #2 (#3)}
\def\PRD#1#2#3{{\sl Phys. Rev.} {\bf D#1}, #2 (#3)}
\def\PLB#1#2#3{{\sl Phys. Lett.} {\bf B#1}, #2 (#3)}
\def\PREP#1#2#3{{\sl Phys. Rep.} {\bf #1}, #2 (#3)}
\def\NPB#1#2#3{{\sl Nucl. Phys.} {\bf B#1}, #2 (#3)}

\def\tev{{\rm TeV }}
\def\gev{{\rm GeV }}
\def\mev{{\rm MeV }}
\def\msusy{M_{\rm SUSY}}
\def\mgut{M_{\rm GUT}}
\def\mz{m_{\rm z}}
\def\mhl{m_{h^0}}
\def\mha{m_{A^0}}
\def\abs#1{\left|#1\right|}
\def\tanb{\tan\beta}

\def\Nbar{{\bar N}}
\def\Hbar{{\bar H}}
\def\sixteenbar{{\overline{16}}}
\def\fivebar{{\overline{5}}}
\def\tenbar{{\overline{10}}}

\def\d{{\rm d}}
\def\refmark#1{ [#1]}

\def\mp{M_{\rm P}}

\def\threebar{{\overline{3}}}
\def\third{\ifmath{{\textstyle{1 \over 3}}}}
\def\half{\ifmath{{\textstyle{1 \over 2}}}}
\def\twothirds{\ifmath{{\textstyle{2 \over 3}}}}
\def\fourthirds{\ifmath{{\textstyle{4 \over 3}}}}
\def\centre#1{{\phantom{.}#1\phantom{.}}}
\def\noteadded{$\underline{\hbox{Note Added:}}$}
\def\etal{{\rm etal.}}
\date={}
\titlepage
\hoffset=-.65cm
\voffset=-.4cm
\hsize=17.5cm
\vsize=23.cm
\baselineskip=24pt
\line{\hfill DESY 94-078}
\line{\hfill MPI-PTh 95-53}
\line{\hfill May 1994}
\line{\hfill hep-ph/9405252}
\line{\hfill Revised: June 1995}
\vskip1.cm

\title{\bf
On the fine-tuning problem in minimal SO(10) SUSY-GUT}

\author{Ralf Hempfling}

\vskip .1in

\centerline{Max-Planck-Institut f\"ur Physik}
\centerline{-Werner-Heisenberg-Institut-}
\centerline{P.O.Box 40 12 12, M\"unchen, Germany}
\vfil
\abstract
In grand unified theories (GUT) based on SO(10) all fermions of one
generation are embedded in a single representation. As a result,
the top quark, the bottom quark, and the $\tau$ lepton have a
universal
Yukawa coupling at the GUT scale. This implies a very large ratio
of
Higgs vacuum expectation values, $\tanb \simeq m_t/m_b$.
We analyze the naturalness of
such a scenario
quantitatively including all the relevant radiative corrections
and find that in minimal unified supergravity
models with universal soft supersymmetry breaking
 parameters at the GUT
scale, the necessary amount of fine-tuning needed is excessive.
GUT threshold correction to the universal
Higgs mass parameter can significantly reduce the
fine-tuning required for such large values of $\tanb$.
We also point out that the top quark mass can play a
crucial role in explaining the hierarchy between
the SUSY breaking scale and the electro-weak scale and, hence,
the naturalness of large values of $\tanb$.


\centerline{
PACS number(s): 11.30.Qc, 12.10.Dm,  12.10.Kt, 12.60.Jv}

\centerline{
to be published in {\sl Physical Review} {\bf D}}
\endpage

\FIG\figcoefficients{%
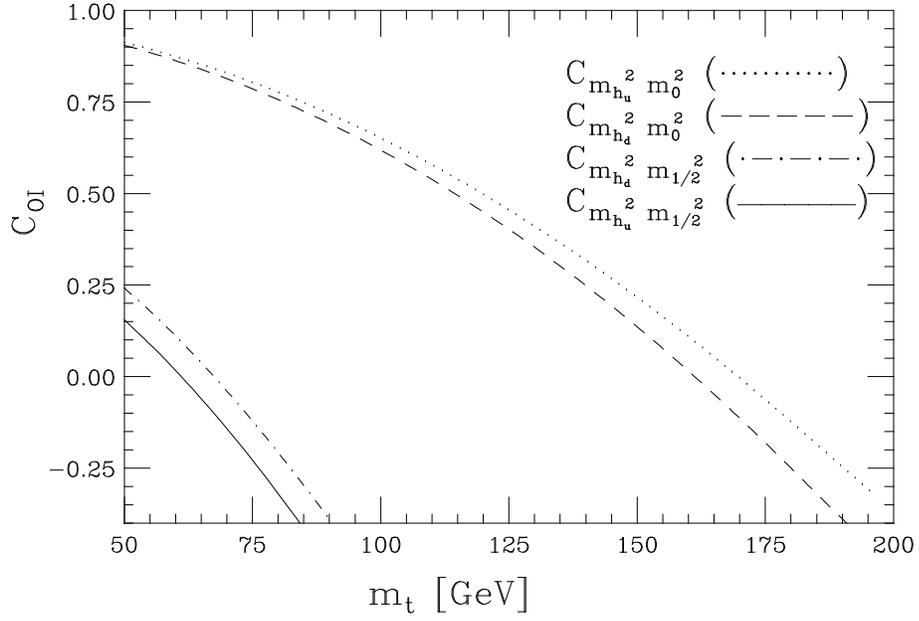
The coefficients $C_{OI}$ ($O = m_{h_u}^2, m_{h_d}^2$ and
$I = m_0^2, m_{1/2}^2$) as a function of $m_t$
}

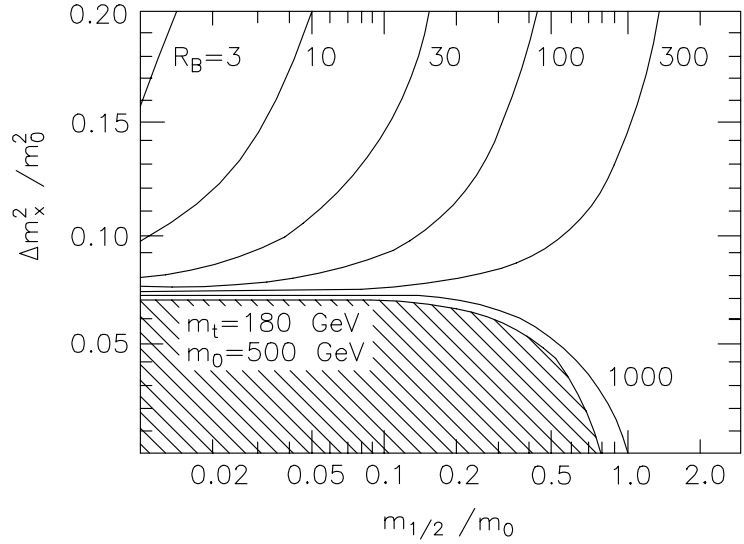
\FIG\figdm{contours of constant $R_B = 3$, 10, 30, 100, 300, 1000
in the $m_{1/2}$--$\Delta m_X^2$ plane (normalized to $m_0$)}

\FIG\figmu{
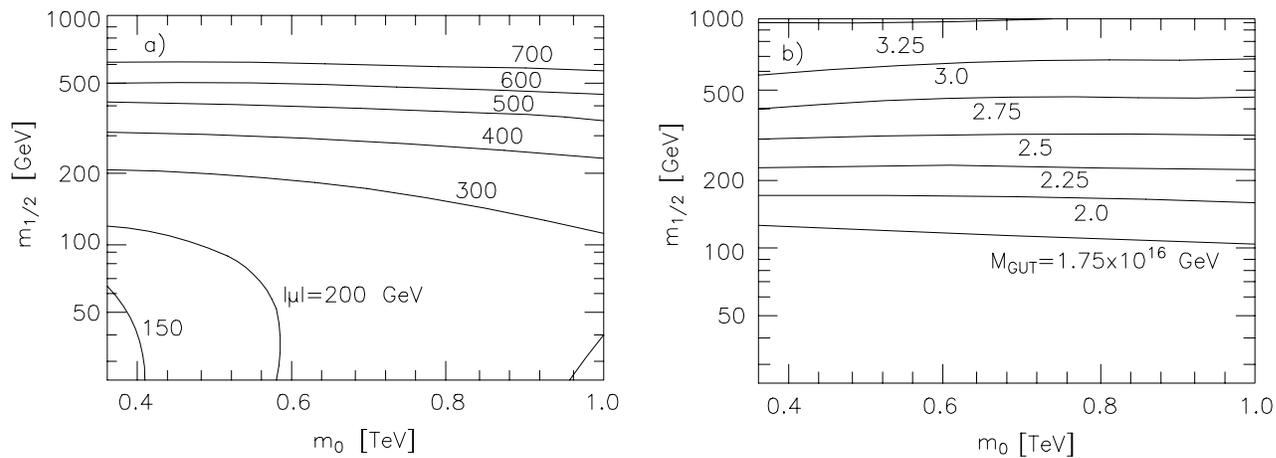
contours of (a) constant $\abs{\mu} = 150$, $200$, $300$, $400$, $500$,
$600$, $700~\gev$ and
(b) constant $\mgut = 1.5$, $1.75$, $2.0$, $2.25$, $2.5$, $2.75$,
$3.0$, $3.25\times 10^{16}~\gev$}

\FIG\figb{
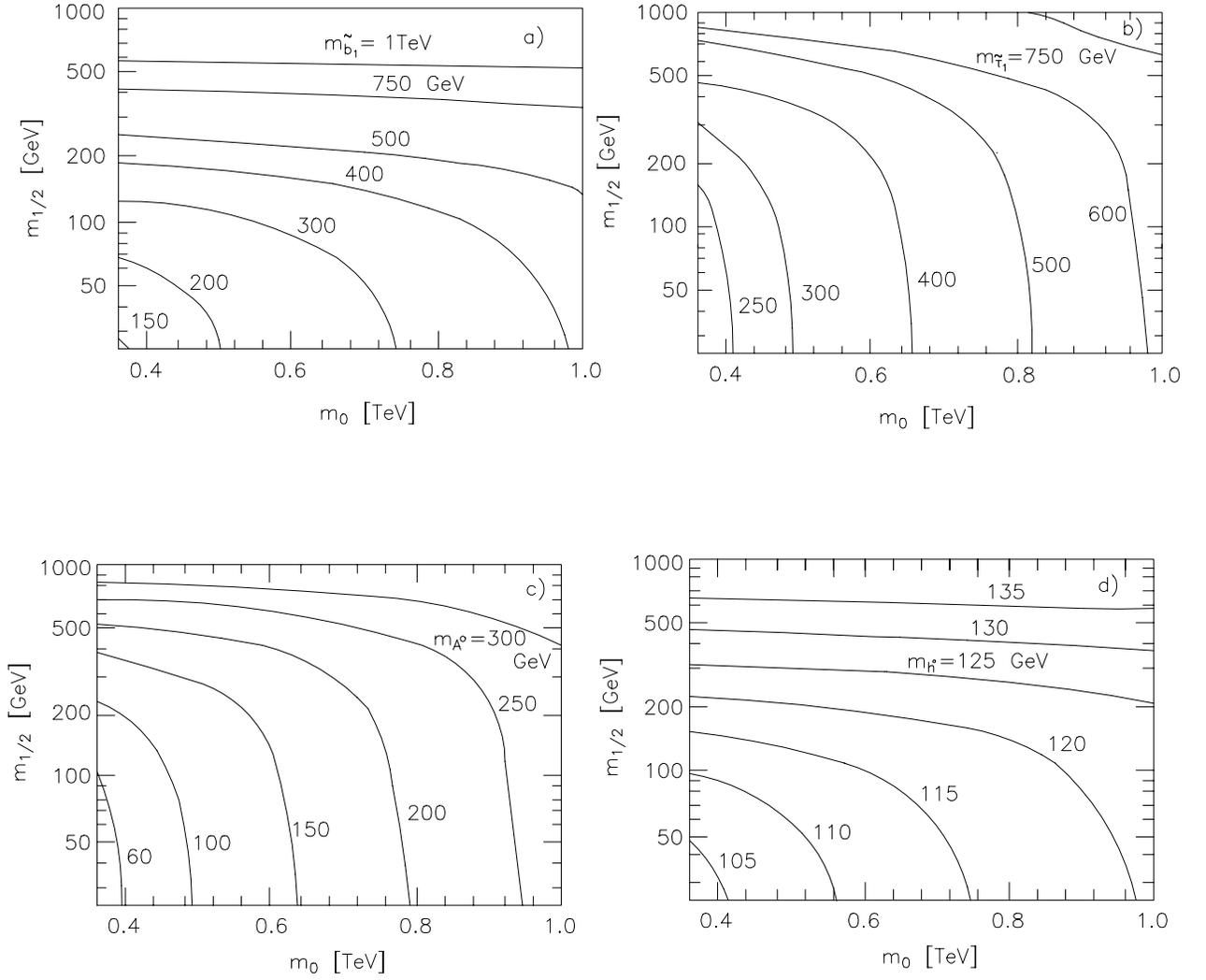
contours of
(a) constant $M_{\tilde b_1} = 150$, 200, 300,
400, 500, $750~\gev$, $1~\tev$,
(b) constant $M_{\tilde \tau_1} = 250$,
 300, 400, 500, 600, $750~\gev$,
(c) constant $\mha =
60$, 100, 150, 200, $300~\gev$ and
(d) constant $\mhl =
105$, 110, 115, 120, 125, 130, $135~\gev$
in the $m_0$--$m_{1/2}$ plane. We set
$A = m_{1/2}$ and $m_t = 180~\gev$}

\FIG\figas{
contours of
(a) constant $\alpha_{\rm s} = 0.126$,
 $0.128$, $0.13$, $0.132$, $0.134$ and
(b) constant pole mass $m_b = 3.8$, $4.0$, $4.25$,
 $4.5$, $4.75$, $5.0$, $5.2~\gev$
in the $m_{1/2}$--$m_0$ plane
for $m_t = 180~\gev$ and $A=m_{1/2}$
}%

\FIG\figfine{
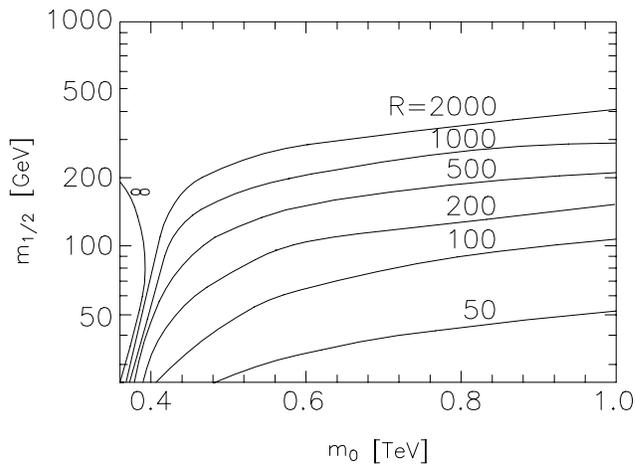
contours of
constant $R = 50$, 100, 200, 500, 1000, 2000
for $m_t = 180~\gev$ and $A=m_{1/2}$}

\FIG\figdas{%
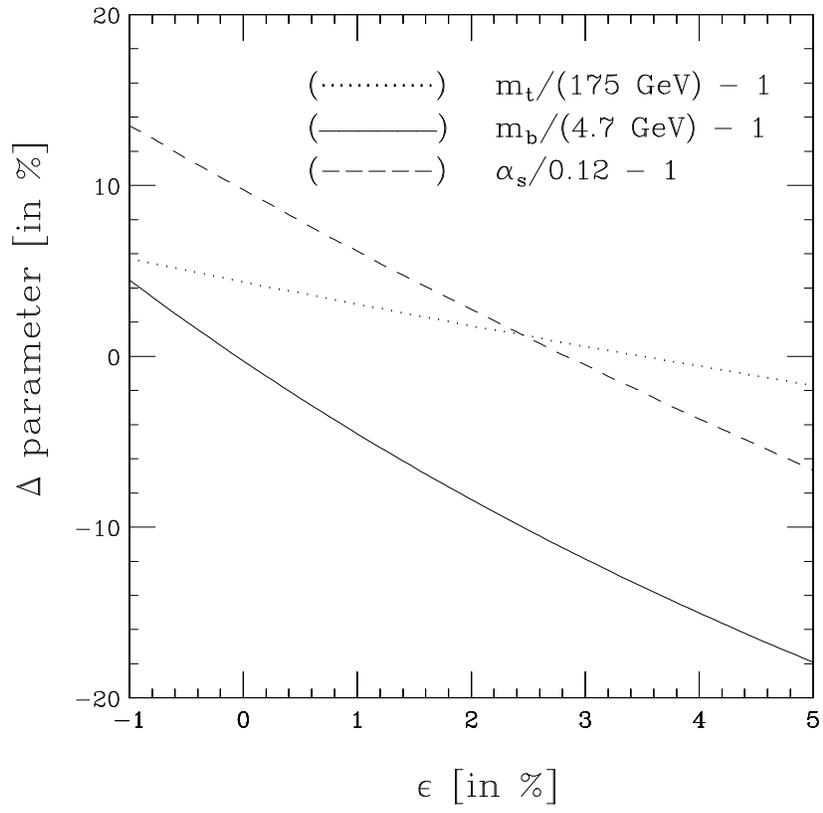
$m_t$, $m_b$ and $\alpha_{\rm s}$ as a function of the
GUT threshold corrections parameterized by $\epsilon$}

\FIG\figdmhmh{%
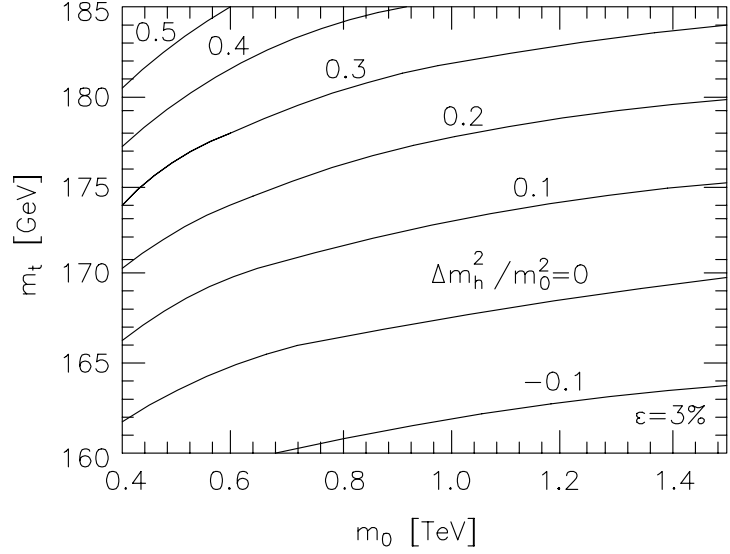
contours of constant $\Delta m_h^2/m_0^2$ in the
 $m_0$--$m_t$ plane.
We fix $m_{1/2} = -\mu = 120~\gev$ for an approximate
$R$ symmetry and $PQ$ symmetry}

\FIG\figdmhfine{%
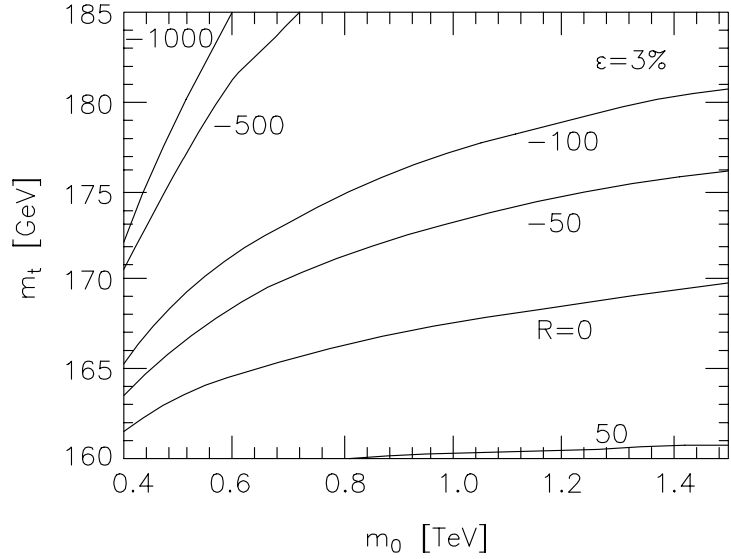
contours of constant $\abs{R} = 0$,
$50$, $100$, $200$, $500$, $1000$, $2000$
in the $m_{1/2}$--$m_0$ plane
for $m_t = 180~\gev$ and $A=0$
}

\FIG\figdmhmb{%
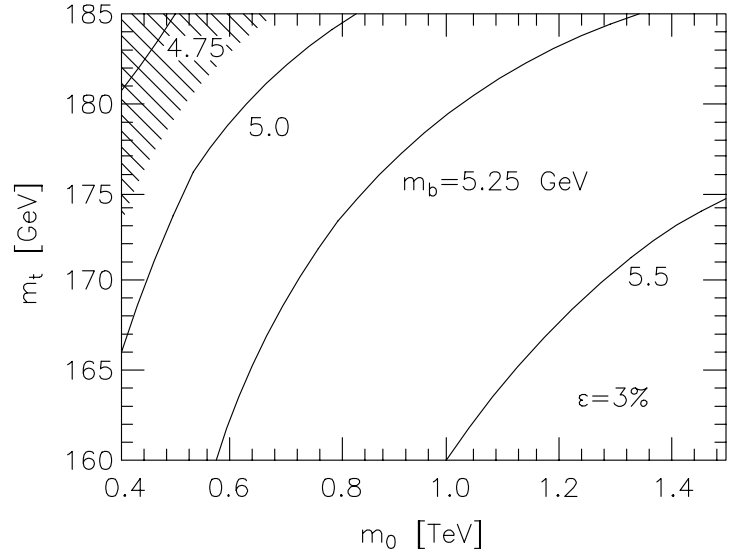
contours of constant pole mass
$m_b = 4.75$, $5.0$, 5.25, $5.5~\gev$
in the $m_0$--$m_t$ plane}

\REF\mssm{for a review, see, \eg, H.P. Nilles,
\PREP {110}{1}{1984};
H.E. Haber and G.L. Kane, \PREP {117}{75}{1985};
R. Barbieri, Riv. Nuovo Cimento {\bf 11}\rm , 1 (1988).}%
\REF\amal{U. Amaldi, W. de Boer and H. F\"urstenau,
 \PLB {260}{443}{1991};
J. Ellis, S. Kelley and D.V. Nanopoulos, \PLB {260}{131}{1991};
P. Langacker and M.X. Lou, \PRD {44}{817}{1992}.}%
\REF\dhr{S. Dimopoulos, L.J. Hall and S. Raby,
 \PRL {68}{1984}{1992};
\PRD {45}{4192}{1992};
V. Barger, M.S. Berger and P. Ohmann, \PRD {47}{1093}{1993};
M. Carena, S. Pokorski and C.E.M. Wagner, \NPB {406}{59}{1993};
P. Langacker and N. Polonsky, \PRD{47}{1093}{1993};
W.A. Bardeen, M. Carena, S. Pokorski and C.E.M. Wagner,
 \PLB{320}{110}{1994}.}%

\noindent{\bf 1. Motivation}

One of the main motivations for supersymmetry (SUSY) is the
cancellation of quadratic divergences in unrenormalized
Green functions. This stabilizes any mass scale under
radiative corrections and allows the existence of
the large hierarchy between the Planck-scale and the electro-weak
 scale.
Furthermore, it has been shown recently that in the minimal
supersymmetric extension of the standard
model (MSSM)\refmark\mssm\ the
SU(3)$_c\otimes {\rm SU(2)}_{\rm L}$
$\otimes {\rm U(1)}_{\rm Y}$
gauge couplings unify at a scale
$\mgut \simeq 2\times 10^{16}~\gev$\refmark\amal.
Additionally, the unification of $\tau$ and bottom Yukawa couplings
at $\mgut$ can be achieved within the MSSM for particular values
 in the
$m_t$--$\tanb$ plane\refmark{\dhr}.
\REF\georgi{%
H. Georgi, in: \sl Proc. American Institute of Physics,
Division of Particles
and Fields, \rm ed. C.E. Carlson (New York, 1975) p. 575;
H. Fritzsch and P. Minkowski,
\sl Ann. Phys. (N.Y.) \bf 93\rm , 193 (1975).}%
Further unification can be achieved if we embed the
SM gauge group within
an SO(10)\refmark\georgi. Here all the fermions of one generation
fit in one single 16 dimensional spinor representation
$F(16)_i$ ($i = 1,2,3$ is the generation index).
Furthermore, SO(10) is anomaly-free and thus provides a nice
explanation
for the cancellation of anomalies within one generation of
standard model
quarks and leptons.

If we assume that the two Higgs doublets, $h_u$ and $h_d$,
required in supersymmetric theories
to give masses to up- and down-type fermions,
are embedded in a single representation, $h(10)$,
and that all the Yukawa couplings of the third generation
come from the unique
coupling $F_3\otimes F_3\otimes h$ we find
$$
y_t(\mgut) = y_b(\mgut)  = y_{\tau}(\mgut)
\,.\eqn\bound
$$
In these models,
the large ratio of $m_t/m_b$ can be explained by
a large ratio of Higgs vacuum expectation values (VEVs)
$$
\tanb \equiv {\VEV{h_u}\over \VEV{h_d}}
\simeq{m_t\over m_b} = {\cal O}(50)\,.\eqn\blabla
$$
\REF\pdg{%
Particle Data Group, L. Montanet \etal, \PRD{50}{1173}{1994}.}%
The prediction for the bottom quark mass from $\tau$-bottom
Yukawa unification tends to be larger than its experimental value,
$m_b = 4.7\pm 0.2~\gev$\refmark\pdg,
unless a significant
part of the QCD correction is cancelled by a large top Yukawa
coupling\refmark\dhr. This predicts that the top quark mass, $m_t$,
must be close to its infra-red (IR) fixed point
which is roughly obtained by setting the $\beta$-function to zero,
\ie
$$\eqalign{%
&{m_t^2\over \sin^2 \beta} + {m_b^2\over 6 \cos^2 \beta}
\simeq (200~\gev)^2
\,.\cr}\eqn\bla$$
\REF\mtexp{%
F. Abe \etal\ [CDF-collaboration], FERMILAB-PUB-95/022-E,
\sl Phys. Rev. Lett.\rm , to be published.}%
\REF\alsh{%
B. Ananthanarayan, G. Lazarides and Q. Shafi,
\PRD{44}{1613}{1991}.}%
\REF\radssb{J. Ellis, J.S. Hagelin, D.V. Nanopoulos
and K. Tamvakis,
\PLB {125}{275}{1983};
G.F. Giudice and G. Ridolfi, \ZPC{41}{447}{1988}.}%
For a top quark mass
of $m_t = 176\pm8({\rm stat})\pm10{\rm (sys)}~\gev$
suggested by recent CDF results\refmark\mtexp\ eq.~\bla\
has two ranges of solutions corresponding to $\tanb \simeq 1$
and to very large values of $\tanb$ that are remarkably
compatible with
eq.~\blabla\ \refmark\alsh.

In minimal unified supergravity (SUGRA) models, where all the
soft SUSY breaking mass parameters are universal, the Higgs
parameters $m_h^2$ ($h = h_u, h_d$) are
driven to negative values by the large Yukawa
couplings\refmark\radssb.
As a result, the correct pattern of
electroweak symmetry breaking occurs naturally in a large
region of the parameter space.
The resulting ratio of Higgs vacuum expectation values can
be expressed in terms of the parameters of the Higgs potential
evaluated at the electroweak scale set by the $Z$
boson mass, $\mz =
91.187$\refmark\pdg.
At tree-level the minimum conditions are
$$
\tan^2\beta = { m_{h_d}^2 +  \mu^2 + \half \mz^2
\over  m_{h_u}^2 + \mu^2 + \half \mz^2}\,.
\eqn\deftanb$$
$$
\sin 2\beta = -{ 2 \widetilde B  \mu\over
 m_{h_d}^2 +  m_{h_u}^2 + 2 \mu^2}
\,.\eqn\defsinb
$$
where $\widetilde B$ denotes the running $B$ parameter evaluated at
the electro-weak scale.
In the large $\tanb$ limit we find that $ \mu$
has to be tuned such that the denominator of eq.~\deftanb\
vanishes as $\sin^2 2\beta$.
The absence of a hierarchy problem then just means that
$\mu^2$ and
$m_{h_u}^2$ should be of the order of or not
much larger than $\mz^2$.
This feature is common to all SUSY models.
\REF\randall{%
A.E. Nelson and L. Randall, \PLB{316}{516}{1993}.}%
\REF\refhrs{%
L.J. Hall, R. Rattazzi and U. Sarid,
\PRD{50}{7048}{1994}.}%
The second condition [eq.~\defsinb] only becomes problematic
in SO(10) type models with large values of $\tanb$\refmark\randall.
It has been pointed out in ref.~\refhrs\
that an approximate symmetry
is the only case where the small values of $\widetilde B\mu$
required in SO(10) by
eq.~\defsinb\ are stable under
radiative corrections and thus natural.
In this paper, we will
combine such a symmetry with the radiative electro-weak
symmetry breaking scenario in minimal SUGRA models and
try to analyse the required fine-tuning quantitatively.

Our paper is organized as follows:
in chapter~2 we review the fine-tuning problem with
universal soft SUSY breaking masses.
In chapter~3 we allow for non-universal soft SUSY breaking
parameters
at $\mgut$ in order to accommodate an approximate symmetry
that yields large
values for $\tanb$ and in chapter~4 we will present our
conclusions.

\noindent{\bf 2. The fine-tuning problem with universal mass
parameters}

In minimal $N= 1$ SUGRA models the effects of SUSY breaking can be
parameterized by only 5 (4 in the MSSM) parameters
$$
{\cal V}_{\rm soft} = \left(A W_3 + B W_2 + C W_1
 + {\rm H.c.}\right)
+m_0^2 \sum_i \phi_i^{\dagger} \phi_i + m_{1/2}
\sum_a \left(\psi_a\psi_a
+{\rm H.c.}\right)\,,\eqn\blabla$$
\REF\pomerol{%
N. Polonsky and A. Pomerol, \PRL{73}{2292}{1994}.}%
where $W_1$, $W_2$ and $W_3$ are the linear, quadratic
and trilinear parts of
the superpotential, respectively (in the MSSM, $W_1 = 0$).
Furthermore, $m_0^2$ ($m_{1/2}$) is the universal mass
parameter of all the
scalar fields $\phi_i$ (gaugino fields $\psi_a$) at the
Planck scale, $\mp$.
In general, the universality of all the soft SUSY breaking
parameters
is violated through radiative corrections
and significant non-universal terms can already be generated
by renormalization group (RG) evolution from $\mp$
to $\mgut$\refmark\pomerol.
In this chapter, we will follow the common practice
of neglecting these model-dependent effects.

\REF\rsymm{%
P. Fayet, \NPB{90}{104}{1975};
A. Salam and J. Strathdee, \NPB{87}{85}{1975}.}%
We will now investigate the feasibility of incorporating
a symmetry that sets
$\mu B = 0$.
Possible candidates for such a symmetry advocated in
ref.~\refhrs~are
the Peccei-Quinn ($PQ$) symmetry
which implies that $\mu = 0$
and $R$ symmetry\refmark\rsymm\
which implies that $B = 0$.
\REF\deffine{%
R. Barbieri and G.F. Giudice, \NPB{306}{63}{1988}.}%
\REF\olechowski{%
M. Olechowski and S. Pokorski, \NPB{404}{590}{1992}.}%
The latter requires also that $A = m_{1/2} = 0$ as can
be seem by inspecting
the $\beta$ function of $B$.
The amount of fine-tuning required to obtain a particular
value of $\beta$,
can be defined as\refmark{\deffine;\olechowski}
$$
R_B = {B\over \sin 2\beta}{\partial \sin 2\beta\over\partial B}
= {B\over \widetilde B}{\partial \widetilde B\over\partial B}
= {B\over \widetilde B}\,.
\eqn\deffine
$$
\REF\massrge{%
K Inoue, A. Kakuto, H. Komatsu and S. Takeshita,
 \PTP{67}{1889}{1982};
J.P. Derendinger and C.A. Savoy, \NPB{253}{285}{1985};
N.K. Falck, \ZPC {30}{247}{1986}.}%
For a natural solution we expect $R_B = {\cal O}(1)$
 and an increasing
$R_B$ means increasing fine-tuning.
Since the set of RG equations for the soft SUSY breaking
mass parameters\refmark\massrge\ is homogeneous
we can write the solutions as
$$
\eqalign{%
& O = C_{OI} I \,,\cr
& O^2 = C_{OI}^2 I^2 \,.}\eqn\blabla
$$
Here, the linear input and output parameters are denoted by
$I = m_{1/2}, A, B$ and $O = m_i$
($i = 1,2,3$), $A_f$ $(f= t,b,\tau$),
$B$, respectively.
The quadratic input and output parameters are denoted by
$I = m_0^2, A^2, A m_{1/2}, m_{1/2}^2$ and $O = m_{\phi}^2$
(where $\phi$ stands for all the scalar multiplets).
The dimensionless coefficients $C_{OI}$ and $C_{OI}^2$
depend only on the coupling constants at $\mgut$ and on
$t \equiv {\rm log}_{10}(\mgut/\msusy)$.
Let us consider the limit of small $\mu$
where the correct electro-weak symmetry
breaking requires that $m_{h_u}^2$
becomes negative.
In fig.~\figcoefficients\ we have plotted the coefficients
$C_{OI}^2$ for $O = m_{h_u}^2, m_{h_d}^2$
and $I = m_{1/2}^2, m_0^2$
as a function of $m_t$.
We see that in the limit $m_{1/2}^2\gg m_0^2$ where
$m_{h}^2 \simeq C_{m_{h}^2 m_{1/2}^2} m_{1/2}^2$ ($h = h_u, h_d$)
the correct spontaneous
electro-weak symmetry breaking can occur already for a top mass of
$m_t \gsim 60~\gev$ assuming $\mu$ is
sufficiently small. We also find
that $m_{h_u}^2 < m_{h_d}^2$ due to the different hypercharges of
the top and bottom quarks
which automatically guarantees that
$\tanb >1$.
On the other hand, in the limit $m_{1/2}^2\ll
 m_0^2$ the spontaneous
electro-weak symmetry breaking can only occur for
$m_t \gsim 165~\gev$. In addition, we see that
$m_{h_u}^2 > m_{h_d}^2$
which yields $\tanb<1$. Thus, this
scenario predicts the wrong hierarchy for the
quark masses of the third generation and is ruled out.

Let us now return to the problem of satisfying
eq.~\defsinb\ without
fine-tuning. By solving the RG equations we obtain
$$\widetilde B \simeq
C_{B B} B + C_{B m_{1/2}} m_{1/2} + C_{B A} A\,,\eqn\solveb$$
where the coefficients are $C_{B B} = 1$,
$C_{B m_{1/2}} \simeq -0.8$ and
$C_{B A} \simeq 0.7$.
Here, we have used
$\alpha_{GUT} = 4.2\times10^{-2}$
for the unified gauge coupling,
$\alpha_{Y} \equiv y_t^2/4\pi = 3\times10^{-2}$
for the top yukawa coupling
(this corresponds roughly to a top quark mass of
$m_t = 180~\gev$; in our numerical
work we shall instead fix $m_t = 180~\gev$ by varying $\alpha_Y$)
and $t = 14$.
By solving eq.~\solveb\ and \deffine\ we obtain
$R_B$ as a function of $A$,
$m_{1/2}$ and $\widetilde B$.

Note that the definition in eq.~\deffine\
is not very satisfying in certain regions of the parameter space.
In particular,
in the region where $m_{1/2} \simeq A$ we find that
$R_B  \simeq 1$ due to a large cancellation
on the right hand side of eq.~\solveb\
($\widetilde B$ being negligible).
However, this does not mean that
no fine-tuning is required along this curve
in the $m_{1/2}-A$ plane
but rather that
it has shifted from tuning $B$ to tuning $A$ and $m_{1/2}$.
Thus, in order to avoid such a cancellation
we simply drop $A$ from eq.~\solveb\ and
obtain a more reliable estimate of the amount of fine-tuning
$$
R_B \simeq {m_{1/2}\over \widetilde B} \,.
\eqn\defrb$$
As we have argued above, we can derive a lower limit for $m_{1/2}$
and thus also for $R_B$
by requiring the correct electro-weak symmetry breaking
[eq.~\deftanb\ and \defsinb].
We obtain the condition
$$\eqalign{%
m_{h_d}^2 - m_{h_u}^2
&={1\over 8\pi^2}\int^{\ln\mgut}_{\ln\mz}
\left(3 y_t^2 X_t-3 y_b^2 X_b
- y_\tau^2 X_{\tau}\right)\d t\cr
&= -0.13 m_0^2 + 0.26 m_{1/2}^2 -0.04 m_{1/2}A
- 0.01 A^2 \gsim \mz^2\,,}
\eqn\dmass
$$
\REF\anant{%
M. Olechowski and S. Pokorski, \PLB{214}{393}{1988};
P.H. Chankowski, \PRD{41}{2877}{1990};
B. Ananthanarayan, G. Lazarides and Q. Shafi,
 \PLB {300}{245}{1993}.}%
where $X_t = m_2^2+m_{\tilde t_R}^2 + m_{\tilde t_L}^2
+A_t^2$ \etc\
Note that
the coefficient of $m_0^2$ in eq.~\blabla\ is
negative in all models where
the top and bottom Yukawa couplings are approximately
 equal due to the
additional effects of the $\tau$ Yukawa coupling.
This can be compensated by a large gaugino mass
parameter\refmark\anant\
which enters with a positive sign.
However, it requires
$m_{1/2} = {\cal O}(m_0)$
which contradicts an approximate $R$ symmetry.

On the other hand, for the $\mu$ parameter we
find from eq.~\defsinb\
for the particular value of $\alpha_Y$ states above
$$
\mu^2
\simeq -\half \mz^2 - m_{h_u}^2
= -\half \mz^2 + 0.06 m_0^2 + 2.35 m_{1/2}^2 -
0.3 m_{1/2} A + 0.07 A^2\,,
\eqn\defmu$$
or $|\mu| \simeq 1.5 m_{1/2}$ which is
inconsistent with an approximate $PQ$ symmetry.
The situation will not change significantly for
a different choice of
 $\alpha_Y$.
Thus, in this minimal version of the model, where all
the mass parameters are universal,
the large value of $\tanb$ can not be protected by
 $R$ or $PQ$ symmetry
can only be obtained by fine-tuning.

\noindent{\bf 3. Non-universal mass parameters}

The simplest extension of the minimal SO(10) SUSY-GUT model
presented above is the inclusion of
non-universal mass parameters.
This entails a loss of predictability but is
certainly favoured over the
introduction of additional fields.
In SO(10) models gauge invariance puts strong
constraints on the form of
these non-universal terms
$$
V_{\rm non-univ} = \sum_{i,j} \Delta m^2_{i j} F_i^{\dagger} F_j
+ \sum_{i,j} \left(A_{i j} F_i F_j h + {\rm H.c.}\right)
+ \Delta m_h^2 h^{\dagger} h\,,
\eqn\defnonuniv$$
\REF\refnewfields{%
L.J. Hall, V.A. Kostelecky and S. Raby,
 \sl Nucl. Phys. \bf B267\rm ,
415 (1986).}%
\REF\refthresh{%
R. Hempfling, \PRD{49}{6168}{1994}.}%
where $\Delta m^2_{i j}$ ($A_{i j}$)
are hermitian (symmetric) and traceless.
[Note, that we have absorbed the Yukawa couplings
into the definition
of $A_{i j}$.]
The terms in eq.~\defnonuniv\
are naturally generated by integrating out additional fields at
$\mgut$\refmark{\refnewfields;\refthresh}
or by RG evolution from $\mp$ to $\mgut$\refmark\pomerol.
\REF\doublettriplet{%
S. Dimopoulos and H. Georgi, \sl Nucl. Phys. \bf B193\rm ,
150 (1981).}%
However, all the terms in eq.~\defnonuniv\
will do very little to solve the basic problem of rendering
the left hand side
of eq.~\dmass\ positive without a large value of $m_{1/2}$.
What we need are different mass parameters for the two Higgs
doublets.
In minimal SO(10) symmetric theories,
where $h_u$ and $h_d$ are assumed to
be embedded in a single representation,
the universality of these two parameters
is based on gauge invariance and cannot be broken explicitly.
Of course, it can be achieved by enlarging the Higgs sector.
However, as we will point out in the next section such an
 extension is
not necessary since the same gauge symmetry that enforces
universality of the
Higgs mass parameter will also break this universality
via the U(1)$_X$ D-term once the SO(10) symmetry is broken.

\noindent{\sl 3.1 Non-universal masses from D-terms}

\REF\solution{%
E. Witten, \sl Phys. Lett. \bf B105\rm , (1981) 267;
L. Ib\'a$\tilde{\rm n}$ez and G.G. Ross,
\sl Phys. Lett. \bf B110\rm , 215 (1982);
P.V. Nanopoulos and K. Tamvakis,
\sl Phys. Lett. \bf B113 \rm , (1982) 151;
A. Masiero, D.V. Nanopoulos, K. Tamvakis and T. Yanagida,
\PLB{115}{380}{1982}.}%

Let us assume for simplicity, that the SO(10)
gauge group is broken to the
SU(3)$_c\otimes {\rm SU(2)}_{\rm L}$
$\otimes {\rm U(1)}_{\rm Y}$
standard model group at $\mgut$
without any intermediate scale and that the breaking
of the U(1)$_X$
symmetry in SO(10)$\supset$SU(5)$\otimes$U(1)$_X$
to SU(5) occurs only via the VEV of $H(16)$ and
$\Hbar(\sixteenbar)$.
This can be achieved \eg\ with the superpotential
$$
W = \lambda S (\Hbar H - m^2) + W_S\,,
\eqn\spot$$
where $m = {\cal O}(\mgut)$, $S$ is a SO(10) singlet and $W_S$
contains the quadratic and trilinear term in $S$.\foot{%
Note, that the superpotential in eq.~\spot\
can only be considered as a simple toy model.
New fields introduced to break SU(5)
and to solve the doublet/triplet
problem\refmark{\doublettriplet;\solution}\
can modify our results quantitatively. However, the
qualitative features of generating a non-vanishing $D$
term will remain.}
In the limit of unbroken $R$ symmetry
(\ie\ $A = B = C = m_{1/2} = 0$)
the potential can be written as $V = V_F + V_D + V_{\rm soft}$
where
$$\eqalign{
&V_F = \sum\abs{{\partial W\over \partial \phi}}^2\,,\cr
&V_D = {g^2\over 8}\abs{\sum \phi^{\dagger}\tau^a\phi}^2\,,\cr
&V_{\rm soft} = \sum m_\phi^2 \phi^{\dagger}\phi\,.}
\eqn\defpot$$
\REF\drees{%
M. Drees, \PLB{181}{279}{1986};
Y. Kawamura, H. Murayama and M. Yamaguchi, \PLB{324}{52}{1994}.}%
Here the sum is over all SO(10) scalar
multiplets $\phi = H,\Hbar,$\etc.
Universality means that $m_\phi^2 = m_0^2$ at $\mp$
for all scalar fields, $\phi$. Note that
gauge-invariance requires that all
fermions of one generation obtain the same mass term at $\mgut$.
With the decomposition
$H = H_Q(10,-1)\oplus H_D(\fivebar,3)\oplus H_N(1,-5)$ and
$\Hbar = \Hbar_Q(\tenbar,1)\oplus \Hbar_D(5,-3)\oplus \Hbar_N(1,5)$
under SO(10)$\supset$SU(5)$\otimes$U(1)$_X$
we can write the minimum of $V$ in the supersymmetric limit
(\ie\ $V_{\rm soft} \equiv 0$) as $\VEV{H_N} = \VEV{\Hbar_N} = m$
with all other VEVs vanishing. If we include $V_{\rm soft}$
then the minimum of $V$ shifts by ${\cal O}(m_0^2/m)$ and
we will find in general that the $D$-term
corresponding to the U(1)$_X$
gauge symmetry no longer vanishes\refmark\drees
\TABLE\particles{The decomposition
of $F$ and $h$ under SU(5)$\otimes$U(1)$_X$
and SU(3)$_c\otimes$SU(2)$_{L}\otimes$U(1)$_Y$}
\bigskip

\vbox{\tenrm
{\narrower\noindent%
{\bf Table~\particles.}\ \ %
The decomposition of $F$ and $h$ under SU(5)$\otimes$U(1)$_X$
and SU(3)$_c\otimes$SU(2)$_{L}\otimes$U(1)$_Y$
\smallskip
}
\tablewidth=17cm
\begintable
  {}  |  \multispan{6}\hfill $F $\hfill
      |  \multispan{4}\hfill $h $\hfill
\cr
SO(10)|  \multispan{6}\hfill $16$\hfill
      |  \multispan{4}\hfill $10$\hfill \crthick
SU(5)$\otimes$U(1)$_X$ |  \multispan{6}\hfill $\quad Q\quad
\oplus\,  D\,  \oplus
\,  N $\hfill
      |  \multispan{4}\hfill $h_U \, \oplus \,  h_D $\hfill
\crthick
SU(5) |  \multispan{3}\hfill $10$\hfill
      |  \multispan{2}\hfill $\fivebar$\hfill
      |  \multispan{1}\hfill $1$\hfill
      |  \multispan{2}\hfill $5$\hfill
      |  \multispan{2}\hfill $\fivebar$\hfill
 \cr
U(1)$_X$ |  \multispan{3}\hfill $-1$\hfill
         |  \multispan{2}\hfill $ 3$\hfill
         |  \multispan{1}\hfill $-5$\hfill
         |  \multispan{2}\hfill $ 2$\hfill
         |  \multispan{2}\hfill $-2$\hfill
\crthick
SU(3)$_c\otimes$SU(2)$_{L}\otimes$U(1)$_Y$
 |  \multispan{3}\hfill $q\, \oplus\,   u^c\, \oplus\,  e^c$\hfill
         |  \multispan{2}\hfill $d^c  \oplus\,  l$\hfill
         |  \multispan{1}\hfill $N$\hfill
         |  \multispan{2}\hfill $t_u\,  \oplus \,  h_u$\hfill
         |  \multispan{2}\hfill $t_d\,  \oplus \,  h_d$\hfill
\crthick
SU(3)$_c$| $3$ | $\threebar$
 |$1$|$\threebar$|$1$|$1$|$3$|$1$|$\threebar$|$1$
\cr
SU(2)$_L$| $2$ |$1$|$1$|$1$|$2$|$1$|$1$|$2$|$1$|$2$\cr
U(1)$_Y$ | $\centre\third$ |$-\fourthirds$|$
\centre2$|$\centre\twothirds$|$-1$
        |$\centre0$|$-\twothirds$|$\centre1$|$
\centre\twothirds$|$-1$
\endtable
}

$$
\VEV{D_X} = {g_X \over 2}\left(X_N \VEV{H_N}^* \VEV{H_N}
             + X_{\Nbar} \VEV{\Hbar_N}^*\VEV{\Hbar_N}\right)
= {1\over g_X X_{N}}\left(m_{\Hbar}^2-m_{H}^2\right)\,,
\eqn\defd$$
where $X_{\Nbar} =-X_N = 5$ are the U(1)$_X$  charges listed in
table~\particles\ with the
normalization $g_X = \sqrt{1/10}g$.
The additional non-universal soft SUSY breaking mass terms
for the Higgs bosons, squarks and
sleptons created by $\VEV{D_X}\neq 0$
can be written as
$$
\Delta V_{\rm soft}
= -\Delta m_X^2\sum_\Phi X_\Phi \Phi^\dagger\Phi\,,
\eqn\mnonuniv$$
where the sum is now over all SU(5)$\otimes$U(1)$_X$  multiplets
listed in table.~\particles.
\REF\eastwood{%
R. Rattazzi,
U. Sarid and L.J. Hall, Stanford Univ. preprint SU-ITP-94-15 and
Rutgers Univ. preprint RU-94-37,
hep-ph/9405313, May 1994,
presented at the 2nd IFT Workshop on Yukawa
Couplings and the Origins of Mass, Feb. 11-13,
1994, Gainsville, FL, USA.}%
The non-universal mass term is given by\refmark\eastwood
$$
\Delta m_X^2\equiv g_X \VEV{D_X}
= {1\over 2 X_{N}}\left(m_{\Hbar}^2-m_{H}^2\right)\,.
\eqn\defd$$
Note that $\Delta m_X^2$ is independent of $g_X$
and the parameters of $W$
indicating that this result is quite general.

Of course, in the limit where
universality is unbroken the right hand side of
eq.~\defd\ vanishes. However, the breaking of universality
at $\mgut$ due to radiative corrections
is almost inevitable.
For example, let us include interaction terms of the form
$$
W\supset \half \left(y_H H H h^\prime
+ y_{\Hbar} \Hbar \Hbar \bar h^\prime\right)\,,
\eqn\newinter
$$
\REF\ioannis{A. Ioannissian, private communication.}%
where $h^\prime \equiv h^\prime(10)$
and $\bar h^\prime \equiv \bar h^\prime(10)$
are different from the Higgs
field $h$.
If we impose
universality ($m_{h^\prime}^2 = m_H^2 = m_\Hbar^2 = m_0^2$)
at $\mp$ and evolve the mass parameters
to $\mgut$ we obtain to first order\refmark\ioannis
$$
m_H^2(\mgut) = m_H^2(\mp) - {10 y_H^2 \over 16\pi^2}
\left(2 m_H^2 + m_{h^\prime}^2\right)
\ln\left({\mp^2\over \mgut^2}\right)+...\,,
\eqn\blabla
$$
where we have omitted all the irrelevant terms.
The non-universal mass term then becomes\foot{%
an interesting alternative to eq.~\newinter\
is $W\supset \sum_i y_i H F_i h^\prime$.
It naturally explains the plus
sign of $\Delta m_X^2$.
Here, one combination of the three
$D_i(\fivebar,3)\subset F_i(16)$
(which should lie dominantly in the first two
generations)
combines with the $h_U^\prime(5,2)
\subset h^\prime(10)$ and becomes massive
while the $h_D^\prime(\fivebar,-2)
 \subset h^\prime(10)$ remains light.
This might also explain the bad mass predictions for the down-type
quarks of the first two generations.}%
$$
{\Delta m_X^2\over m_0^2} = {y_{ H}^2 - y_{\bar H}^2 \over 2 X_N}
\times {\cal O}(3)\,.
\eqn\blabla$$
Finally, by running the mass parameters to $\msusy$ we find
to a good approximation
$$\Delta V_{\rm soft}
\simeq -\Delta m_X^2 \sum_i\left(X_{\phi_i}-Y_{\phi_i}\right)
\phi_i^{\dagger}\phi_i\,,\eqn\dvsoft$$
where the sum is over all MSSM scalar particles
with the charges listed in table~\particles.
In particular, we find
$$m_{h_d}^2 - m_{h_u}^2 \simeq 2 \Delta m_X^2+...\,,
\eqn\blabla
$$
where the ellipse stands for the contributions of the
universal terms [eq.~\dmass].

\noindent{\sl 3.2 $R$ symmetry}

We now come to a numerical evaluation of the fine-tuning
in the minimal SO(10) model including non-zero values
of $\Delta m_X^2$. First, we note that
for $\Delta m_X^2/m_0^2 \simeq 7\%$
the right hand side of eq.~\dmass\ becomes
positive independent of the value
of $m_{1/2}$.
\REF\fcnc{F. Gabiani and A. Masiero,
 \sl Nucl. Phys. \bf B322\rm , (1989);
J.S. Hagelin, S. Kelley and T. Tanaka, \NPB{415}{293}{1994}.}%
By inspecting the RGEs we find that the
squark mass parameters for the first
two generations are numerically irrelevant.
The same is true for the
flavor changing neutral current (FCNC)
mediating off-diagonal entries once
we impose the experimental constraints\refmark\fcnc.
Thus, the only important scalar mass parameters are
sfermion mass parameter of the third generation
and the Higgs mass parameter which we take to be
universal for now (\ie\ we set $\Delta m_{i j}^2
= \Delta m_{h}^2 = 0$).
The top-quark mass (or alternatively $\alpha_Y$) can be fixed
by imposing exact $\tau$-bottom Yukawa unification.
However, there are still significant theoretical uncertainties.
This becomes manifest if we try to unify the
down-type quark and lepton
Yukawa couplings of the first two generations.
Here, one finds that the predicted
value for $(m_d m_\mu)/(m_s m_e) \simeq 1$
differs from the experimental
value\refmark\pdg\ by a factor of order 10.
The prediction for $m_b$ from Yukawa unification
in a minimal model is, therefore, not
expected to be reliable to more
than about the mass of the strange quark
$m_s \lsim 300~\mev$\refmark\pdg
which translates into a huge incertainty in the prediction of
$\alpha_Y$.
Thus, we allow values of $m_t$ to vary within its
experimental bounds\refmark\mtexp\ (in all plots where $m_t$ is fixed
we chose $m_t = 180~\gev$).

In fig.~\figdm\ we present contours of constant fine-tuning,
parameterized by $R_{B} = 3$, 10, 30, 100, 300, 1000
in the $m_{1/2}$--$\Delta m_X^2$ plane normalized to
$m_0 = 500~\gev$ for $m_t=180~\gev$
and $A = 0$.
The shaded area is ruled out by requiring that
$\tanb>1$.
Acceptable values of $\tanb$ for $\Delta m_X^2 = 0$
can only be found for
values of $m_{1/2} \simeq 0.7 m_0$\refmark\anant\
where the fine-tuning is excessive and
exceeds the lower bound of $R_B \gsim \tanb$
derived in ref.~\randall\ by more than an order of magnitude.
However, in the region where $\Delta m_X^2/m_0^2\gsim 7\%$
the value of $m_{1/2}$ and thus also the value of $R_B$
can become much smaller.

We now investigate whether the region of acceptable fine-tuning is
compatible with the phenomenological constraints.
\REF\pdecay{%
R. Arnowitt, A.H. Chamseddine, and P. Nath, \PLB{156}{215}{1985};
P. Nath, A.H. Chamseddine, and R. Arnowitt, \PRD{32}{2348}{1985};
T.C. Yuan, \PRD{33}{1894}{1986};
P. Nath and R. Arnowitt, \PLB{287}{89}{1992}.}%
\REF\francesca{%
S. Bertolini, F. Borzumati, A. Masiero and G. Ridolfi,
\NPB{353}{591}{1991};
F. Borzumati, \ZPC{63}{291}{1994}
and references therein.}%
\REF\gordy{%
G.L. Kane, C. Kolda, L. Roszkowski and J.D.
 Wells, \PRD{49}{6173}{1994};
\PRD{50}{3489}{1994}.}%
Here, we only impose the constraints coming
 from direct particle searches
in order to keep our results as general as possible.
In particular, we do not impose any constraints coming from
the proton lifetime\refmark\pdecay,
the decay $b\to s\gamma$\refmark\francesca, or
relic abundance\refmark\gordy.

The most important constraint can be derived from
non-observation of the decay of the $Z$ boson into a chargino pair
($Z\rightarrow \chi_i^+ \chi_j^-$, $i,j = 1,2$)
at LEP\refmark\pdg.
It implies that $m_{1/2} \gsim 50~\gev$ which is similar
to the bound on $m_{1/2}$ from the decay
$Z\rightarrow \chi_i^0 \chi_j^0$, $i,j = 1,2,3,4$.
This constraint alone eliminates
almost the entire region in fig.~\figdm\
where $R_B \lsim 30$.
No significant constraints arise from the non-observation of
sfermions since this
only constrains the region of small $m_0^2$
(for fixed $m_{1/2}$)
where $R_B$ is large anyways.
Thus, we expect to be able to decrease $R_B$ by increasing $m_0$.
However, in the region where $m_0 \gg \mz$ we have to take into
account the fine-tuning
required to satisfy eq.~\deftanb.
In the large $\tanb$ limit we find that $\mu$
 has to be tuned such that
the denominator in eq.~\deftanb\ vanishes as sin$^2 2\beta$.
The corresponding fine-tuning
is given by
$$R_\mu \equiv
{\mu^2\over m_{h^0}^2}{\partial m_{h^0}^2 \over \partial \mu^2}
={\mu^2\over m_{h^0}^2}{\partial m_{h^0}^2 \over \partial \mu^2}
= 2{\mu^2\over m_{h^0}^2}\,,
\eqn\deffine$$
\REF\refdmhl{%
Y. Okada, M. Yamaguchi and T. Yanagida,
{\sl Prog. Theor. Phys.} {\bf 85}, 1 (1991);
H.E. Haber and R. Hempfling, \PRL{66}{1815}{1991};
J. Ellis, G. Ridolfi and F. Zwirner, {\sl Phys. Lett.}
{\bf B257}, 83 (1991).}%
\REF\reflep{%
D. Decamp \etal, \PREP{216}{253}{1992}.}%
where we have substituted $\mhl^2$ for $\mz^2$.
This choice is more appropriate since eq.~\deftanb\
is derived from minimizing the Higgs potential.
At tree-level these quantities are equal up
 to terms of the order of
$1/\tan^2\beta$
but large radiative corrections can increase
 $\mhl^2$ by a factor of
up to two\refmark\refdmhl.
Since we see from eq.~\defmu\ $\mu$ has to grow with $m_0$
(with a coefficient that depends on $\alpha_Y$)
the absence of fine-tuning defines an upper limit of $m_0$.
\REF\langacker{%
P. Langacker and N. Polonsky,
 \PRD {47}{4029}{1993}; \PRD{49}{1454}{1994}.}%

We now proceed to evaluating the combined fine-tuning,
parameterized by $R \equiv R_B
R_\mu$ numerically. Four of the nine GUT
input parameters, $A$, $B$, $\mu$, $m_{1/2}$,
$m_0^2$, $\Delta m^2_X$, $\alpha_{\rm GUT}$, $\alpha_{y}$
and $\mgut$,
are fixed by experimental observables\refmark{\pdg;\langacker}
$$\eqalign{
&\sin^2\theta_{\rm w} = 0.231\,,\cr
&\alpha_{\rm em}^{-1} = 127.9\,,\cr
&\mz = 91.187~\gev\,,\cr
&m_{\tau} = 1.7771~\gev\,.}\eqn\blabla
$$
\REF\topdown{%
for more details about the approach as well as the
treatment of the threshold
corrections see: R. Hempfling, \ZPC{63}{309}{1994}.}%
We fix the non-universal term $\Delta m_X^2/m_0^2 = 0.1$
whose sole purpose it is to assure that $m_{h_u}^2<m_{h_d}^2$ but
which is otherwise of minor importance.
Furthermore, we chose $m_t = 180~\gev$,
and $A = m_{1/2}$ and we present the results in the
$m_{1/2}$--$m_0$
plane in the ranges $380~\gev \lsim m_0 \lsim 1~\tev$
and $25~\gev \lsim m_{1/2} \lsim 1~\tev$.

Once these parameters are specified we can compute the full
supersymmetric particle spectrum.
We define $\mgut$ as the scale where SU(2)$_L\otimes$U(1)$_Y$
coupling constants intersect. Here, we include
 all leading logarithmic
SUSY threshold corrections by decoupling the superpartners
from the RGEs at scales below their masses as well as
the finite thresholds enhanced by a factor of
 $\tanb$\refmark\topdown.
The $\mu$ parameter is fixed by imposing the minimum conditions
[eq.~\deftanb\ and \defsinb].
In fig.~\figmu\ we present contours of
(a) constant $\abs{\mu} = 150~\gev$, $200~\gev$,
 $300~\gev$, $400~\gev$, $500~\gev$,
$600~\gev$, $700~\gev$ and
(b) constant $\mgut = 1.75$, $2.0$, $2.25$, $2.5$, $2.75$,
$3.0$, $3.25\times 10^{16}~\gev$. We find that
$\abs{\mu} = O(m_{1/2})$
unless $m_{1/2}\lsim\mz$ where we have $\abs{\mu} = O(m_0/3)$

\REF\hoang{%
R. Hempfling and A. Hoang, \PLB{331}{99}{1994}.}%
The lightest squarks and sleptons are expected to be
the ones of the third generation due to the
effects of the Yukawa couplings.
In fig.~\figdm\ we present contours of
(a) constant lightest bottom squark mass, $M_{\tilde b_1}$,
(b) constant lightest $\tau$ slepton mass, $M_{\tilde \tau_1}$,
(c) constant CP-odd Higgs mass, $\mha$, and
(d) constant lightest CP-even Higgs mass, $\mhl$
in the $m_0$--$m_{1/2}$ plane.
Here, $M_{\tilde \tau_1}$, $M_{\tilde b_1}$
and $m_{A^0}$ are evaluated at
tree-level and the prediction for $\mhl$
includes one-loop and two-loop radiative
corrections\refmark{\refdmhl;\hoang}.
We see that in all of the parameter space under consideration
the particle spectrum is heavier
than the experimental limits from direct
particle search at LEP or TEVATRON\refmark\pdg.

\REF\runtopole{%
H. Arason \etal, \PRD{46}{1992}{3945};
N. Gray, D.J. Broadhurst, W. Grafe
and K. Schilcher, \ZPC{48}{1990}{673}.}%
 From unification of gauge and Yukawa couplings we can also predict
$\alpha_{\rm s}$ and $m_b$.
In fig.~\figas\ we present contours of
(a) constant $\alpha_{\rm s} = 0.126$,
$0.128$, $0.13$, $0.132$, $0.134$ and
(b) constant pole mass $m_b = 3.8$, $4.0$, $4.25$,
 $4.5$, $4.75$, $5.0$, $5.2~\gev$
in the $m_{1/2}$--$m_0$ plane.
The values of $\alpha_s$ is about one standard deviation above the
world average value $\alpha_s(\mz) = 0.12 \pm 0.01$\refmark\pdg.
This discrepancy which is still within the experimental errors
might also be indication for physical effects such as GUT threshold
corrections.
In order to prevent these possible effects
to propagate into our prediction for $m_b$
we use the experimental value
$\alpha_{\rm s}(\mz) = 0.112$ from deep inelastic
scattering\refmark\pdg\foot{%
we favor this value over LEP data since it is extracted at
an energy scale at which the dominant contributions to
$m_b$ arise.}%
for the evolution of $m_b$ from $\mz$ to
$m_b$ and to convert the running
mass into the pole mass\refmark\runtopole\
using two-loop threshold corrections and two-loop $\beta$
functions.
We find that the so obtained
prediction of $m_b$ is somewhat smaller than experiment
for large values of $m_{1/2}$ due
to large SUSY threshold corrections to
$m_b$\refmark{\refhrs;\refthresh;\topdown}.
The cusps in the contours of fig.~\figas~(a) and (b)
for $m_{1/2} \simeq 100~\gev$
correspond to the decoupling of the wino
and might be an artifact of our
leading logarithmic approximation of the threshold corrections.
This behaviour is expected to change slightly if one uses
full one-loop threshold corrections.
However, we do not expect any significant
effects from non-logarithmic terms.

Finally, in fig.~\figfine\ we present contours of
constant $R = 50$, $100$, $200$, $500$, $1000$, $2000$
in the $m_{1/2}$--$m_0$ plane.
We see that the situation has improved
significantly with respect to
the universal case where only the region
where $m_{1/2}\gsim 0.7 m_0$
\ie\ $R = O(10^3)$, is allowed.
Nonetheless, a lower limit of $R \gsim 50$ remains if we impose
the experimental constraint $m_{1/2} \gsim 50~\gev$.

\noindent{\sl 3.3 $PQ$ symmetry}

In section 3.2 we have investigated the scenario where large
values of $\tanb$ are protected from radiative corrections by an
approximate $R$ symmetry.
However, this symmetry together with the experimental lower limits
on the chargino masses require a heavy sfermion spectrum
[\ie, $m_0 = {\cal O}(\tanb m_{1/2}$)]
which in turn
re-introduces a hierarchy problem.
The same problem arises if we instead
try to impose an approximate $PQ$ symmetry.
\REF\refrpq{%
M. Dine and D. MacIntire, \PRD{46}{2594}{1992};
R. Hempfling, \PLB{329}{222}{1994}.}%
The best option is therefore to impose $R$ symmetry and
$PQ$ symmetry simultaneously.
Such a correlation between $R$ symmetry and $PQ$ symmetry
is quite natural in models where the $\mu$ parameter is
generated dynamically\refmark\refrpq.
The correct electro-weak symmetry breaking can
be achieved in this case
by further relaxing the universality constraints.
Namely, we can fix $\mu$ to any arbitrary value
by allowing for non-zero values for $\Delta m_h^2$.
In this case we have to tune $\Delta m_h^2$ instead of $\mu$ in
order to obtain the correct value for the VEVs.
Thus, the fine-tuning parameter defined in eq.~\deffine\
has to be replaced by
$$
R_\mu \rightarrow R_{\Delta m_h^2} = {\Delta m_h^2 \over \mhl^2}
{\partial \mhl^2 \over \partial \Delta m_h^2}
\simeq 1.2 {\Delta m_h^2 \over \mhl^2}
\,,\eqn\defrdmh
$$
For our numerical analysis we will chose $m_{1/2} =
 -\mu = 120~\gev$
which yields a sufficiently heavy chargino/neutralino
mass spectrum
(the sign of $\mu$ is chosen such that the SUSY
threshold corrections
to $m_b$ are negative).

\REF\gutthreshold{%
R. Barbieri and L.J. Hall, \PRL{68}{752}{1992}.}%
Furthermore, in fig~\figas\ (a) we have seen
that the prediction for $\alpha_{\rm s}$
tends to be somewhat larger than the experimental value.
Thus, let us also include the effects of GUT threshold corrections.
These corrections can be included
by modifying the boundary conditions at
$\mgut$
$$
\alpha_1 = \alpha_2 = \alpha_3 (1+ \epsilon)\,,\eqn\blabla
$$
where all the corrections are absorbed in
the single parameter $\epsilon$.
In fig.~\figdas\ we present the prediction of
$\alpha_{\rm s}$ (dashed curve), $m_t$
(dotted curve) and $m_b$ (solid curve)
as a function of $\epsilon$.
We see that for $\epsilon = 1\%$ the
prediction of $\alpha_{\rm s}$, $m_b$ and
$m_t$
for fixed $\alpha_{\rm GUT}$ and $\alpha_Y$
decrease by about $3\%$, $3\%$ and $1\%$, respectively.
The magnification of the threshold corrections to $\alpha_{\rm s}$
by a factor of $\alpha_{\rm s}/\alpha_{\rm GUT}$
 can be understood easily.
On the other hand, the corrections to $m_b$ and $m_t$ are of
higher order and more complicated. Nonetheless, their dependence
on $\epsilon$
is quite strong.

In any realistic model these threshold corrections
are indeed expected to be significant\refmark\gutthreshold.
Thus, rather than imposing exact unification at $\mgut$
we determine the threshold correction, $\epsilon$
by fixing $\alpha_{\rm s} = 0.12$ \ie\ we set $\epsilon = 3\%$.
We can now use the value of $\alpha_{\rm s}$
obtained from unification for the evolution of $m_b$ also at scales
below $\mz$ rather than using experimental results
as we have done in
fig.~\figas~(b).
Furthermore, we vary $m_t$ within its
phenomenologically allowed range.

In fig.~\figdmhmh\
we present contours of constant $\Delta m_h^2/m_0^2$ in the
$m_0$--$m_t$ plane.
The contour $\Delta m_h^2 = 0$ corresponds to the case
of universal Higgs and squark masses.
We see that in the limit of approximate $R$ symmetry
and $PQ$ symmetry (in the present case: $\mu/m_0, m_{1/2}/m_0
=  0.4 \sim 0.08$) and with $\Delta m_X^2$ being the only non-universal
soft SUSY breaking parameter the correct electro-weak
 symmetry breaking can
only be achieved for $m_t = 162\sim 170~\gev$.
In turn, we can say that if the top quark mass falls
within this range
and the Higgs and top squark mass parameter are universal
we naturally obtain
a large hierarchy $m_0^2/\mz^2 \ll 1$
in the limit of approximate $R$ symmetry
and $PQ$ symmetry.
This point is illustrated in fig.~\figdmhfine\
where we present contours of constant
$R$ in the $m_t$--$m_0$ plane and
find that for the range of $m_t$ denoted above
 the fine-tuning is quite small.
Note, the existence of a contour $R = 0$ in fig.~\figdmhfine\ which
is equivalent to the contour $\Delta m_h^2 = 0$
in fig.~\figdmhmh\ is an unphysical artifact of our
definition of the fine-tuning in eq.~\defrdmh.
For small values of $\Delta m_h^2$
it is more appropriate to replace $\Delta m_h^2$ by
the average deviation from universality,
\ie\
$$
\Delta m_h^2 \to \sqrt{\half \left[\left(\Delta m_h^2\right)^2
+\left(\Delta m_h^2\right)^2\right]}\,.\eqn\blabla
$$
With this replacement the fine-tuning parameter $R$ would remain
roughly constant within the contours $|\Delta m_h^2/m_0^2| =
\Delta m_X^2/m_0^2 = 1/10$.
It is interesting to note that
these contours correspond roughly to $|R| = 50 \simeq \tanb$,
a limit that has already been established by more
general considerations in ref.~\randall.
Note also that $R$ is largely
independent of $m_0$ but exhibits a
strong dependence on $m_t$.

In fig.~\figdmhmb\ we present contours of constant pole mass
$m_b$ in the $m_t$--$m_0$ plane.
Here, the strong coupling constant lies in the range
$\alpha_{\rm s} = 0.120\pm 0.002$.
We find that the predicted value of $m_b$ lies
outside the experimental
error bars (shaded area). The reason for the difference between
this plot and  fig.~\figas\ (b)
is twofold. First we have used a larger value of $\alpha_s$ from
$\mz$ to $m_b$.
Second, by imposing $R$ symmetry and $PQ$ symmetry we have severely
constrained the size of the
SUSY threshold corrections to $m_b$ which are proportional to
$\mu m_{\tilde g}/m_0^2\ll 1$.
In particular, we see
that without any GUT corrections for the Yukawa
couplings there is no overlap between the area of small fine-tuning
(say $\abs{R} \simeq 50$) and the area where $m_b$
falls within the experimental error.
However, as we have pointed out earlier
additional (possibly non-renormalizable) terms
have to be introduced to explain the masses
of the first two generations.
If we keep in mind that in minimal SO(10)
the universal Yukawa coupling determined by the charm quark mass
$m_c = 1.2~\gev$ can only account
for a small fraction of the strange quark mass
then we can expect the theoretical uncertainty in $m_b$
to be of the order of
$m_s \lsim 300~\mev$\refmark\pdg.
Thus, by allowing the entire region where
the predicted values of $m_b$ in fig.~\figdmhmb\
is below $5.2~\gev$ there
will be a region in parameter space with
$m_t \simeq 165~\gev$ where
the fine-tuning is of the order of $\tanb$.

\noindent{\bf 4. Conclusions}

We have presented a detailed analysis of the
naturalness of the large
$\tanb$ region required in minimal SO(10) SUSY-GUT models.

We confirm that in the case of universal soft
SUSY breaking parameters
the particle spectrum is dominated by a large gluino mass
and the fine-tuning needed for the correct electroweak
symmetry breaking
is ${\cal O}(10^3)$ and excessive.

The situation improves if we include a non-zero U(1)$_X$ $D$-term.
This opens a whole new range in parameter space where
$R$ symmetry is approximately conserved
and the required fine-tuning is only ${\cal O} (\tanb)$.
The region of smallest fine-tuning is characterized by a
light chargino
and neutralino just above the experimental bounds.

Finally, we use non-universal soft SUSY breaking parameters for
the Higgs sector
and the sfermion sector in order to impose an additional
approximate $PQ$ symmetry.
We find that in this scenario the top quark mass is the
crucial parameter.
In particular, for $m_t = O(165~\gev)$ almost universal
 boundary conditions
will result in a hierarchy between
the electroweak scale, $\mz$, and the SUSY scale, $m_0$.
This can further improve the fine-tuning situation in the
large $\tanb$ region.

This scenario is characterized by a $m_t$ significantly below the
its IR fixed point. As a result, the
prediction of $m_b$ is above the experimental value.
However, we argue that the resulting prediction falls still within
the significant theoretical errors
related to GUT scale threshold correction and to
the origin of the Yukawa couplings for the first two generations.

\ack{I would like to thank A. Ioannissian, M. Carena, C. Wagner, S. Pokorski,
R. Rattazzi and U. Sarid for many pleasant and useful discussions.}
\REF\cpow{%
M. Olechowski and S. Pokorski, \PLB{344}{201}{1995}.}%
\REF\nilles{%
D. Matalliotakis and H.P. Nilles, \NPB{435}{115}{1995}.}%

\noindent\noteadded{ During the final stages of this work I became aware of
other work on related
matters\refmark{\eastwood;\cpow;\nilles} where similar conclusions were
obtained.
\refout
\figout
\end